\documentclass[%
 reprint,
 longbibliography,
floatfix,
 amsmath,amssymb,
 aip,
  jcp,
 superscriptaddress
]{revtex4-1}

\usepackage{dcolumn}
\usepackage{bm}

\usepackage[normalem]{ulem}

\newcommand{\tx}[1]{\text{\tiny #1}}
\newcommand{\tb}[1]{_\tx{#1}}
\newcommand{\tu}[1]{^\tx{#1}}

\newcommand{\dd}{\text{d}}
\newcommand{\kB}{k_\text{B}}

\newcommand{\EQ}[1]{\begin{align}#1\end{align}}

\newcount\markedup
\ifnum\markedup=1

\newcommand{\new}[1]{{\color{blue} #1}}
\newcommand{\newx}[1]{{\color{magenta} #1}}
\else
\newcommand{\new}[1]{{\color{black} #1}}
\newcommand{\newx}[1]{{\color{black} #1}}
\fi

   \usepackage[scaled=0.95]{helvet}
 \usepackage{tgtermes}

\usepackage{xcolor}
\definecolor{nblue}{RGB}{0,74,153}
\definecolor{negg}{RGB}{255,255,255}

\usepackage[hyperfootnotes=false]{hyperref}	 
\hypersetup{
  colorlinks,
  citecolor=nblue,
  linkcolor=nblue,
  urlcolor=nblue
}

\usepackage[nameinlink]{cleveref}

 \crefname{equation}{{\color{black}{Eq.}}}{{\color{black}{Eqs.}}}
 \creflabelformat{equation}{#2\textup{(#1)}#3}
 \crefname{figure}{Fig.}{Figs.}
 \Crefname{figure}{Figs.}{Figs.} 
 \crefname{table}{Table}{Tables}
 \crefname{appendix}{{\color{black}{Sec.}}}{{\color{black}{Secs.}}}
 \crefname{section}{{\color{black}{Sec.}}}{{\color{black}{Secs.}}}

\usepackage{graphicx}
\graphicspath{{figures/}}
 
 \usepackage{caption}
\usepackage{ragged2e}
 \DeclareCaptionJustification{justified}{\justifying}
\usepackage{siunitx}

\let\originalleft\left
\let\originalright\right
\renewcommand{\left}{\mathopen{}\mathclose\bgroup\originalleft}
\renewcommand{\right}{\aftergroup\egroup\originalright}

\begin{document}


\title{Tracer dynamics in polymer networks: generalized Langevin description
}

\author{Sebastian Milster}%
\email{sebastian.milster@physik.uni-freiburg.de}
\affiliation{Physikalisches Institut, Albert-Ludwigs-Universität Freiburg, Hermann-Herder Strasse 3,
D-79104 Freiburg, Germany}
 
\author{Fabian Koch}
\affiliation{Physikalisches Institut, Albert-Ludwigs-Universität Freiburg, Hermann-Herder Strasse 3,
D-79104 Freiburg, Germany}

\author{Christoph Widder}
\affiliation{Physikalisches Institut, Albert-Ludwigs-Universität Freiburg, Hermann-Herder Strasse 3,
D-79104 Freiburg, Germany}

\author{Tanja Schilling}
\email{tanja.schilling@physik.uni-freiburg.de}
\affiliation{Physikalisches Institut, Albert-Ludwigs-Universität Freiburg, Hermann-Herder Strasse 3,
D-79104 Freiburg, Germany}

\author{Joachim Dzubiella}
\email{joachim.dzubiella@physik.uni-freiburg.de}
\affiliation{Physikalisches Institut, Albert-Ludwigs-Universität Freiburg, Hermann-Herder Strasse 3,
D-79104 Freiburg, Germany}
\affiliation{Cluster of Excellence livMatS @ FIT – Freiburg Center for Interactive Materials and Bioinspired Technologies, Albert-Ludwigs-Universität Freiburg, D-79110 Freiburg, Germany}

\date{\today}

\begin{abstract}
Tracer diffusion in polymer networks and hydrogels is relevant in biology and technology, while it also constitutes an interesting model process for the dynamics of molecules in fluctuating, heterogeneous soft matter.  Here, we study systematically the time-dependent dynamics and (non-Markovian) memory effects of tracers in polymer networks based on (Markovian) implicit-solvent Langevin simulations. In particular, we consider spherical tracer solutes at high dilution in regular, tetrafunctional bead-spring polymer networks, and control the tracer--network Lennard-Jones (LJ) interactions and the polymer density. Based on the analysis of the memory (friction) kernels, we recover the expected long-time transport coefficients, and demonstrate how the short-time tracer dynamics, polymer fluctuations, and the viscoelastic response are interlinked.  Further, we fit the characteristic memory modes of the tracers with damped harmonic oscillations and identify LJ contributions, bond vibrations, and slow network relaxations, which enter the kernel with an almost linear scaling with the LJ attractions. This procedure proposes a reduced functional form for the tracer memory, allowing for a convenient inter- and extrapolation of the memory kernels. This leads eventually to highly efficient simulations utilizing the generalized Langevin equation (GLE), in which the polymer network acts as an additional thermal bath with tuneable intensity.

\end{abstract}


\maketitle

\section{Introduction}

The diffusive transport of molecules through dense polymer networks, such as hydrogels, is a key process for many technological applications, e.g., for filtration\cite{Shannon2008Mar,Geise2010Aug,Pendergast2011Jun} and nanocatalysis,\cite{Renggli2011Apr,Roa2017Sep,carregal2010catalysis,Nanoreactor} while it has also high functional relevance in biomolecular systems\cite{Broedersz2014Jul, hydrogel,Metzler:mucin} and in biomedical applications,\cite{biomedical} such as drug-delivery,\cite{Li2016Oct,Brudno2015Dec,Moncho-Jorda2020Nov,Mitchell2021Feb}  tissue engineering,\cite{tissue,Stamatialis2008Feb,Lee2018Dec} soft robotics,\cite{softrobotics} or responsive microreactors.\cite{Bell2021Jan,Milster2023Oct} Apart from acting as selective diffusion barriers for long time transport,\cite{C6NR09736G, LIELEG2011543} the hydrogel matrix may regulate the diffusive molecular encounter and thus (bio-) chemical signaling and reactions also on microscopically short timescales in the crowded matrix.\cite{Burla2020Feb,10.3389/fphy.2020.00134,Pete,WEISS2014383}  A better understanding and control of these often anomalous diffusive processes are thus desired for the design and optimization of soft hydrogel-based (biomimetic) functional materials. 

Hence, a large body of works has contributed in the last decades to the understanding of the microscopic mechanism behind single molecule (`tracer') dynamics and transport in polymer matrices on different scales.\cite{kusumi2005paradigm,Franosch2011Oct, Ernst2014, Fany,WOLDEKIDAN2021463, 10.3389/fphy.2020.00134,  Fytas, Roichman}  In particular, improved experimental single particle tracking methods, for example, by fluorescence microscopy, have elucidated the stochastic motion and complex physical nature of tracer motion on a wide range of microscopic timescales with high resolution. The goal of these experiments, as well as the numerous accompanying theoretical works, is to learn about the dynamical features of both, the underlying medium as well as the coupled motion of the tracer particles in these usually crowded, disordered and highly fluctuating environments.

Occurring phenomena, like anomalous (sub-) diffusion, non-Markovianity, and memory, are often intertwined processes, and far from being well understood. Especially in polymeric systems, such as melts and networks, slow relaxation, memory and subdiffusion are commonplace,~\cite{Kremer1990Apr, Groot1995Aug, Kopf1997Nov, Kavanagh1998Jan, Watanabe1999Nov,Gurtovenko2000Aug, Gurtovenko2005Jul,Panja2010Jun, Vandoolaeghe2005Jul,Rizzi2020Jul,vegt2021} and can connect to the tracer dynamics for strong coupling.~\cite{Fytas, Holm, Singh2020} However, most of the theoretical studies on tracer diffusion in polymer networks have  focused on the long-time dynamics, e.g., long-time self-diffusion or static friction.\cite{Masaro1999Aug,Amsden1998Nov,Kamerlin_2016, Kim2020,WOLDEKIDAN2021463,hansing, Cai, Schweizer, Milster2021,Queseda} It was demonstrated that such dynamics sensitively depends on microscopic details, such as polymer density, mesh-size (or crosslink density), and in particular the tracer--polymer interactions,~\cite{Fytas} defining effective energy landscapes for diffusion.~\cite{PhysRevLett.122.108001, Kim2020,Milster2021}

\begin{figure*}\centering
 \includegraphics[width=\linewidth]{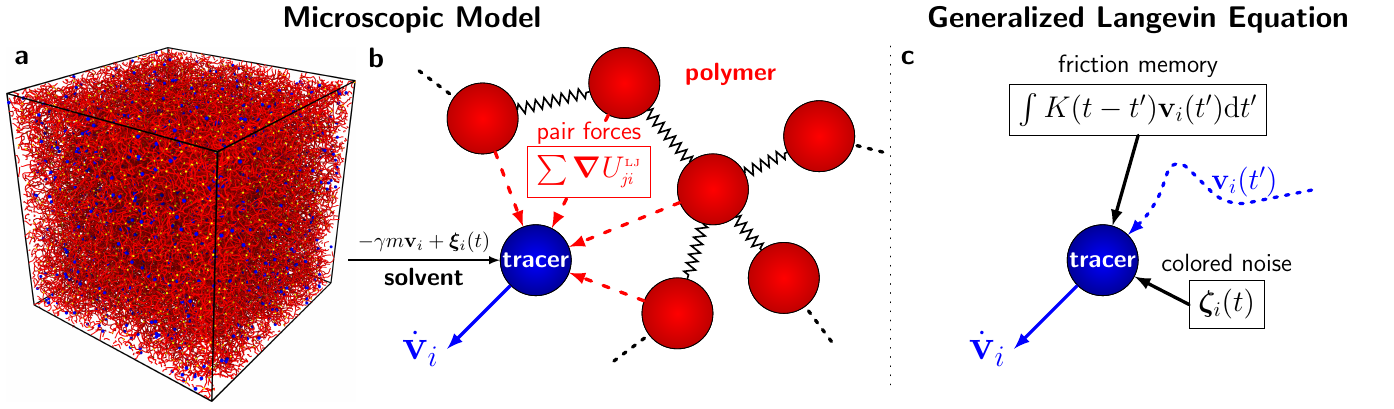}
 \caption{\label{fig:snapshot} 
 {\bfseries (a)}: Simulation snapshot of the microscopic (Markovian, implicit-solvent) model of diffusive tracer particles (blue) in a (periodic) polymeric bead-spring network (red: chain segments, yellow: crosslinkers). While the tracer particles are displayed with correct size ($\sigma$), the chain monomer and crosslinker diameters are reduced for illustration.  {\bfseries (b)}: Schematic representation of the microscopic model displaying classical Langevin solvent contributions and pairwise deterministic forces acting on the tracer particles. {\bfseries (c)} Coarse-grained generalized Langevin equation (GLE) model that incorporates solvent and polymer bath effects through the memory kernel $K$ and correlated fluctuations $\zeta$. }
\end{figure*}

While the long-time dynamics has been well studied, we are now interested in the physics of the full timescale dependent tracer dynamics in a fluctuating polymer network, and how it is governed (`slaved') by the physical coupling to the polymer network dynamics through interactions. Intuitively, repulsive tracers in dilute networks should be weakly coupled to the polymer fluctuations, while more attracted tracers in dense networks are expected to pick up much more of the dynamic features of the matrix.  How the short-time tracer dynamics is controlled by the microscopic interactions in detail is still challenging to categorize.~\cite{Fytas}

A popular tool to model and interpret anomalous dynamics in (macro-) molecular liquids revolves around memory (friction) kernels in the framework of the non-Markovian generalized Langevin equation (GLE),\cite{zwanzig1961memory,mori1965transport,Kubo1966,Shugard1977Mar, Shin2010Oct, goychuk2012viscoelastic, jung2017iterative, netz2019,Lesnicki2016Apr, Meyer_2020_Non, Meyer_2020_Numerical,Straube2020Jul, vegt2021,Doerries2021Mar} leading, for example, to recent GLE descriptions for anomalous polymer~\cite{Panja2010Jun,panja2010generalized,Tian2022Dec} and tracer dynamics.~\cite{Singh2020} The often {\it a priori} assumed functional forms of the tracer kernel reflect directly the polymer matrix fluctuations in a strong tracer--polymer coupling limit,\cite{Rizzi2020Jul,Grebenkov2013Oct,goychuk2012viscoelastic} applicable, e.g., 
for larger probes (larger than mesh size) in the framework of microrheology.\cite{Winter1987Dec, Waigh2005Feb,Tassieri2012Nov, Robertson-Anderson2018Aug}
Nevertheless, a amenable quantification of the coupled tracer--polymer dynamics for different interaction strengths on the full time-scale is still lacking, and this work aims to fill this gap.

Our strategy and outline is as follows: \new{In \cref{sec:micromodel} we start  from a relatively simple model of a regular bead-spring network consisting of crosslinked monomer-resolved beads interacting with spherical tracer particles. We model the  solvent implicitly by employing a Markovian Langevin thermostat to the molecular dynamics simulation. From such a `microscopic' model, we then dynamically coarse-grain to the level of the tracer particles by integrating out the degrees of freedom of the polymeric medium as theoretically outlined in \cref{sec:GLE,sec:ACF,sec:computeK,sec:diffusion_methods,sec:GSER}. Particularly, we calculate and analyze the appropriate tracer and polymer memory (friction) kernels in the time and frequency domain.}

\new{In \cref{sec:ACFana,sec:kernelana} we examine systematically the velocity and force autocorrelations as well as the kernels' behavior under the influence of the changing tracer--network interactions, and for different densities. We further present  various consistency checks for the long-time dynamics based on the autocorrelations and the memory in \cref{sec:longtimediff}. We discuss in detail how the time-dependent friction depends on the coupling to the polymer dynamics (\cref{sec:network_dyna,sec:freq_domain}), where we clearly observe large correlations for larger attractions (a few times the thermal energy).}

\new{In \cref{sec:dampedharmonics} we finally demonstrate that only a few modes of the kernel are necessary for a massively reduced GLE description of the tracer dynamics, where the magnitude of the friction modes can be described by simple, almost linear scaling laws as a function of the tracer--polymer interaction strength. This enables efficient predictions of the tracer dynamics in different simulation setups in which the polymer plays the role of an additional while tuneable heat bath.
}

\section{Model and Methods}
\subsection{Microscopic model}\label{sec:micromodel}

In our microscopic simulations, we consider spherical solute molecules (tracers) diffusing inside a polymeric medium, where the latter is modeled by a bead-spring network similar to previous works,\cite{Kim2020,Milster2021,Q:nonlinear} and as visualized in \cref{fig:snapshot}(a). The polymer network consist of tetrafunctional crosslinkers connecting the polymer chains (ten monomers long) in a regular fashion.

The temporal evolution of the $i$th particle's velocity is governed by the Markovian Langevin equation, reading
\EQ{
m\dot{\mathbf{v}}_i(t) = \mathbf{F}\tu{int}_i(t) -\gamma_0 m\mathbf{v}_i(t)+\boldsymbol{\xi}_i(t),\label{eq:CLE}
}
where the deterministic forces $\mathbf{F}\tu{int}_i(t)$ result from bonded and non-bonded particle interactions, and where $-\gamma_0 m \mathbf{v}_i(t) + \boldsymbol{\xi}_i(t)$ are the implicit solvent's contributions, namely a viscous drag, $-\gamma_0 m \mathbf{v}_i(t)$, and a \newx{Gaussian} white noise force with zero mean, $\langle\boldsymbol{\xi}_i(t)\rangle=0$, and intensity $\langle{\xi}_{i\nu}(t){\xi}_{j\nu'}(t)\rangle=2\gamma_0 m k\tb{B} T\delta(t-t')\delta_{ij}\delta_{\nu\nu'}$, where  $\nu,\nu'\in\{x,y,z\}$ are the Cartesian coordinates. Here, $\gamma_0$ denotes the friction coefficient in the absence of any other polymer and tracer particles, i.e.\ for $\mathbf{F}\tu{int}_i(t)=0$ $\forall t$. It determines the free equilibrium diffusion via the Einstein relation, $D_{0}=\kB T / (m \gamma_0)$, where $k\tb{B}T=\beta^{-1}$ is the thermal energy, and it also defines the Brownian timescale $\tau_0=\gamma_0^{-1}$, which we use as reference.

All (non-bonded) particles interact via Lennard-Jones potentials,
\EQ{
 U_{ij}\tu{LJ}(r_{ij})=4\varepsilon_{ij}\left[\left(\frac{\sigma}{r_{ij}}\right)^{12}-\left(\frac{\sigma}{r_{ij}}\right)^{6}\right],
}
with distance $r_{ij}=\left|\mathbf{r}_i-\mathbf{r}_j\right|$, and particle diameter $\sigma$, which we chose identical for the network and the tracer beads. The potential strength $\varepsilon_{ij}$ depends on the particle pair, resulting in network--network ($\varepsilon\tb{nn}$), tracer--tracer ($\varepsilon\tb{tt}$), and tracer--network ($\varepsilon\tb{tn}$) interactions.

 The polymer bonds are modelled by harmonic springs,
 \new{\EQ{ 
 U\tb{b}(r_{ij})=\kappa\tb{b}\left(r_{ij}-r_0\right)^2,
 }}
 with strength $\kappa\tb{b}=100~k_B T/\sigma^2$  and length \new{$r_0=\sigma$}. The angle potentials of the network are flexible springs, described by
 \EQ{  
 U\tb{a}(\Psi_{ijk})=\kappa\tb{a}\left(\Psi_{ijk}-\Psi_{ijk}\tu{0}\right)^2,}
 with strength $\kappa\tb{a}=1~k\tb{B} T/\text{rad}^2$, and equilibrium angle $\Psi\tu{0}\tb{mer}=\pi$ or $\Psi\tu{0}\tb{x}=\arccos(-1/3)$ (tetrahedral angle), dependent on the vertex particle type $j$, which can be either chain monomer (`mer') or crosslinker  (`x').
 
 We fix the bonded network parameters and employ different network--network interaction strengths, $\beta\varepsilon\tb{nn}\in\left\{0.35, 0.5\right\}$, resulting in typical polymer volume fractions of roughly $\phi\in\left\{0.16,0.32\right\}$ after $NpT$ equilibration.\cite{carregal2010catalysis,Kim2020,Milster2021}
 
 The tracer particles interact mutually repulsively, i.e., the tracer--tracer interaction strength is fixed, $\beta\varepsilon_{tt}=0.1$. \new{However, we choose a low density, $c\approx0.005\sigma^{-3}$, such that the forces $\mathbf{F}\tu{int}_i$ acting on a single tracer are known to be practically only governed by the interaction with the polymer network.\cite{Kim2020,Milster2021}  More details on the microscopic simulations and a tabulated summary of all parameters are provided in the Appendix \labelcref{app:mic_sim}.}

\subsection{Generalized Langevin equation}\label{sec:GLE}
In this work, we reduce the high-dimensional Markovian model for the tracer dynamics by coarse-graining the polymer network (cf. \cref{fig:snapshot}). The Markovian Langevin equation [\cref{eq:CLE}] can be then replaced by a non-Markovian integro-differential equation, namely the generalized Langevin equation (GLE), reading
\EQ{
 m\dot{\mathbf{v}}_i(t) = -m\int\limits_{0}^{t}\mathrm{d}t' K(t-t') {\mathbf{v}}_i(t')+\boldsymbol{\zeta}_i(t),\label{eq:GLE}
}
where the memory kernel, $K(t-t')$, and the correlated fluctuations $\boldsymbol{\zeta}_i(t)$, comprise all polymer and implicit solvent bath effects of the underlying model. \new{In fact, $K$ carries also some tracer--tracer cross-correlations, which, in principle, can be corrected.\cite{Klippenstein2021May,Klippenstein2022Jul} Nonetheless, with the choice of a relatively low tracer concentration ($c=0.005\sigma^{-3}$), we may regard $K$ as a single-tracer memory. A comparison with a test simulation with an even lower tracer concentration ($c\approx0.001\sigma^{-3}$, not presented in this work) suggests that the effect of the tracer--tracer interactions on the correlations and the memory is of the order of a few percent.}

\new{In \cref{eq:GLE}, the components} of the fluctuating force $\zeta_{i\nu}(t)$ with $\nu\in\{x,y,z\}$ and the memory kernel $K(t-t')$ are related via the fluctuation-dissipation theorem (FDT),
\EQ{
	\langle{\zeta}_{i\nu}(t){\zeta}_{j\nu'}(t')\rangle &=K(t-t')m\kB T \delta_{ij}\delta_{\nu\nu'}.
}
Note that the kernel $K(\tau)$ in the present work does not change over time and depends only on the delay $\tau=t-t'$ (with arbitrary $t$ and $t'$).  If the memory decays practically instantaneously, $K(\tau)\to\gamma\delta(\tau)$, we recover the structure of the Markovian Langevin equation with friction $-m\gamma\mathbf{v}_i(t)$ and delta-correlated fluctuations. More details on the derivation of the GLE and the formal definitions of the quantities appearing in the GLE are summarized in the Appendix \labelcref{appendix_GLE}.

\subsection{Relation to the autocorrelation function}\label{sec:ACF}
By multiplying the GLE [\cref{eq:GLE}] with $\mathbf{v}_i(0)$ and taking the ensemble average, we obtain the equation of motion for the autocorrelation dynamics
\EQ{\label{eq:eom_autocorrelation}
\frac{\dd}{\dd t} C_v(t) &= - \int\limits_0^t\dd t' K(t-t') C_v(t').
}
Here, the orthogonality of the fluctuating force and the initial velocity is used, namely $\langle\zeta_{i\nu}(t)v_{i\nu}(0)\rangle=0$, and we introduce the velocity autocorrelation function (VACF)
\EQ{
	C_v(\tau)=\frac{1}{3}\langle \mathbf{v}(\tau)\mathbf{v}(0)\rangle.
}
The prefactor results from averaging over the Cartesian coordinates due to the isotropy of the system. Hence, at zero delay it takes the value $C_v(0)=\kB T/m$.\\
By taking the Laplace transform of \cref{eq:eom_autocorrelation},
\begin{align}\label{eq:volterra_s}
	s\hat{C}_v(s)-C_v(0) &= -\hat{K}(s)\hat{C}_v(s),
\end{align}
and using the 1D Green-Kubo relation for the diffusion constant $D$, we find 
\begin{align}\label{eq:green_kubo_relation}
	D &=\int\limits_0^\infty\dd\tau\,C_v(\tau)=\hat{C}_v(0)=\frac{C_v(0)}{\hat{K}(0)}.
\end{align}
Employing the Einstein relation, we obtain a relation between the effective friction constant and the integral over the memory kernel,
\begin{align}\label{eq:einstein_relation}
	\gamma = \frac{\kB T}{mD} = \hat{K}(0).
\end{align}
This friction constant is compatible with the correct diffusion constant by definition.\\
Further, one can use the relation between Laplace transforms and one-sided Fourier transforms, $\tilde{\mathcal{F}}(\omega) = \hat{\mathcal{F}}(-i\omega)$, to define the complex diffusivity
\EQ{\label{eq:Domega}
	\tilde{D}(\omega)&:=\hat{C}_v(-i\omega)=\int\limits_{0}^\infty\mathrm{d}\tau C_{v}(\tau) e^{i\omega\tau}.
}
The complex diffusivity, which is proportional to the \emph{generalized} dynamic mobility $\tilde{\mu}(\omega)$, \cite{Henery1971Sep,Kubo1966,Balakrishnan2021}  is related to the kernel via [c.f. \cref{eq:volterra_s}]
\EQ{
	\tilde{D}(\omega)=\kB T \tilde{\mu}(\omega)=\frac{\kB T}{m (-i\omega+\hat{K}(-i\omega))}.\label{eq:CVKrelation0}
}
The one-sided Fourier transform of the memory kernel, $\tilde{K}(\omega) = \hat{K}(-i\omega)$, provides the frequency spectrum $|\tilde{K}(\omega)|$ of the fluctuating forces $\zeta_{i\nu}(t)$ in \cref{eq:GLE},\cite{Groot1985} and can also be interpreted as the frequency-dependent friction entering the viscoelastic response, which we discuss later.

In our analysis of the microscopic system we will also evaluate its force autocorrelation function (FACF),  defined as
\EQ{
	C_F(\tau)=\frac{1}{3}\langle \mathbf{F}(\tau)\mathbf{F}(0)\rangle,
}
with $\mathbf{F}(t)=m\dot{\mathbf{v}}(t)$ from the microscopic velocities, which is related to the VACF via $C_F(\tau)=-m^2\ddot{C}_v(\tau)$.

\subsection{Calculation of the memory kernel\label{sec:computeK}}
To calculate the memory kernel we use the equation of motion for the autocorrelation dynamics, \cref{eq:eom_autocorrelation}, which does not involve the fluctuating force anymore. In principle, if one knows the autocorrelation function of the observable, which can easily be calculated from simulation data, \cref{eq:eom_autocorrelation} can be used to obtain the memory kernel. As Mori already showed,~\cite{mori1965transport} formally this equation can be solved via a Laplace transform. \cref{eq:CVKrelation0} allows, in principle, to compute $\tilde{K}(\omega)$ from the VACF in the frequency domain,\cite{vegt2021,netz2019} however, due to numerical subtleties and possible artefacts of the Fourier treatment,\cite{netz2019} we calculate the kernel in the time domain. In the latter the memory kernel is usually extracted via an iterative or recursion ansatz. \cite{Berkowitz_1981_Memory,Berne_1970_Calculation,jung2017iterative,Meyer_2020_Non} In this article we use a modified version of the matrix-inversion method introduced in ref.~\citenum{Meyer_2020_Numerical} adapted to the stationary case. The details of the adaptations are given in the Appendix \labelcref{appendix_numerics_of_memory_kernel}.\\
However, some properties of the memory kernel can be inferred directly. Calculating the time derivative of \cref{eq:eom_autocorrelation},
\begin{align}\label{appeq:CFCV}
	\frac{C_F(t)}{m^2} &= K(t)C_v(0) + \int\limits_0^t\dd\tau K(\tau)\frac{\dd}{\dd t}C_v(t-\tau),
\end{align}
we see that the memory kernel carries some characteristics of the correlation functions. At short correlation times we find \new{ $K(\tau)\approx\left(\beta/m\right)~C_F(\tau)$}, and from \cref{eq:CVKrelation0} we see that in the low frequency limit, when $|\omega|\ll |\tilde{K}(\omega)|$, we recover the long time Einstein relation, $\tilde{K}(\omega)=[\beta m \tilde{D}(\omega)]^{-1}$.

\subsection{Effective long-time transport coefficients\label{sec:diffusion_methods}}
In this section we briefly recall common methods to calculate the effective long-term equilibrium tracer diffusion, $D$, and friction, $\gamma$, which are connected via the Einstein relation [first equality in \cref{eq:einstein_relation}].

A very common approach to calculate $D$ from simulation trajectories and single-particle tracking experiments is the use of the mean-squared-displacement (MSD),\cite{BarratHansen}
\EQ{
D\tb{MSD}=\lim_{t\to\infty}\frac{\langle(\mathbf{r}(t)-\mathbf{r}(0))^2\rangle}{6 t}.}


Equivalently to the MSD method, the diffusion coefficient can be computed from the integral of the velocity autocorrelation function [first equality in \cref{eq:green_kubo_relation}], \cite{Kubo1966} coinciding with the zero-frequency limit of \cref{eq:Domega}, i.e., $D\tb{VACF}=\tilde{D}(0)$.

As shown in \cref{sec:ACF}, this is also consistent with calculating the friction constant via the integral over the memory kernel \cref{eq:einstein_relation}. From the FDT we also know that the same result could be obtained by integrating over the autocorrelation of the fluctuating force
\begin{align}
	\gamma &= \frac{1}{\kB T m} \int\limits_0^\infty\dd\tau\,C_{\zeta}(\tau).
\end{align}
However, here it is important to note that this relation holds for the fluctuating forces $\zeta_{i\nu}(t)$ from the projected dynamics and not the total forces $F_{i\nu}(t)=m\dot{v}_{i\nu}$ appearing in the experiment or simulation. The two following expressions are not the same:
\begin{align}
	F_{i\nu}(t) &= m \exp(t\mathcal{L})\mathcal{L}v_{i\nu},\\
	\zeta_{i\nu}(t) &= m\exp(t\mathcal{Q}_\text{M}\mathcal{L})\mathcal{Q}_\text{M}\mathcal{L}v_{i\nu}.
\end{align}
Here, $\mathcal{L}$ is the Liouvillian and $\mathcal{Q}_\text{M}$ is the orthogonal Mori projection operator (c.f. Appendix \labelcref{appendix_GLE}). The fluctuating forces from the projected dynamics are usually much more difficult to obtain and, hence, the autocorrelation of the forces from the experiment or simulation are often used to estimate friction constants.
Only for a tracer particle with infinite mass \new{or under specific constraints}, the Green-Kubo relation for FACF yields the correct friction coefficient,\new{\cite{Espanol1993force,vogelsang1987determination,kirkwood1947statistical,netz2017,Bocquet1994Jul,espanol2019solution,Hijon_2010_Mori,Akkermans_2000_Coarse}} ${\beta}\int_0^\infty\mathrm{d}\tau \lim_{m\to\infty} C_F(\tau)=m\gamma\tb{FACF}$.
Finite masses, however, \new{are not covered by the Green-Kubo relation and} 
lead to the \emph{plateau problem} and the integral vanishes. Hence, the integral
\EQ{
\gamma\tb{FACF}=\frac{\beta}{m}\int\limits_0^{\tau\tb{cutoff}} \mathrm{d}\tau C_F(\tau)
}
needs to be cut off at $\tau\tb{cutoff}\approx\gamma^{-1}$, which depends on the effective friction itself. In practice, cutting off the integral at its maximum provides results comparable to the other methods. We assume that some confusion regarding the plateau problem has been caused in the past by the fact that $\zeta_{i\nu}(t)$ is called a fluctuating force in the literature on projection operator formalisms, while $F_{i\nu}(t)$ is a force, which fluctuates.

\def\sc{1.3}
\begin{figure}
 \begin{tabular}{c}
\includegraphics[scale=\sc]{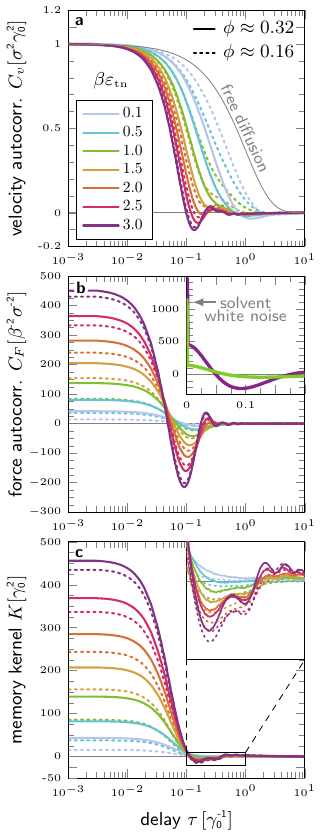} 
 \end{tabular}

 \caption{\label{fig:CvCfK}Velocity autocorrelation of the tracer, $C_{v}$, [panel (a)], force autocorrelation, $C_{F}$, [panel (b)], and memory kernel, $K$, [panel(c)], vs. delay time $\tau$ for different tracer--polymer interaction strengths, $\beta\varepsilon\tb{tn}\in\{0.1, ... ,3.0\}$, [see color code in panel (a)], and two different polymer volume fractions, $\phi\in\{0.16, 0.32\}$, drawn as dashed and solid lines, respectively. Gray zero value lines are added for orientation. The inset in panel (b) depicts two FACFs ($\phi\approx0.32,\beta\varepsilon\in\{1.0,3.0\}$) for small times with linear axis scaling to highlight the solvent-related delta peak at $\tau=0$. Note that the same delta peak rescaled by $\beta/m$ appears in the memory kernel but is not visible due to the logarithmic axis scaling. The inset in panel (c) magnifies the region $\tau\in[0.1,1]$, $K\in[-20,10]$.}
\end{figure}

\subsection{Generalized Stokes-Einstein relation}\label{sec:GSER}
The Fourier transform of the memory kernel $\tilde{K}(\omega)$ plays an important role under mild nonequilibrium conditions, e.g., in the presence of weak sinusoidal driving $\tilde{f}\tb{ext}(\omega)$, because it provides information about the elastic and viscous contributions. In the frequency domain, $\tilde{K}(\omega)$ directly enters the mobility or, equivalently, the  mechanical (viscoelastic) impedance, reading $\tilde{\mu}^{-1}=m({-i\omega+\tilde{K}})$, and one can already conclude that the viscous part, $\text{Re}(\tilde{K})$, is related to the (frequency-dependent) dissipation.\cite{ChaikinLubensky, BarratHansen}  The imaginary part $\text{Im}(\tilde{\mu}^{-1})$ contains inertial effects $m\omega$, and elastic interactions with the surrounding stored in $\text{Im}(\tilde{K})$.

In the linear-response regime, the \emph{generalized} Stokes-Einstein relation (GSER) provides the connection to the dynamic viscosity $\tilde{\mu}^{-1}={3\pi\sigma}\tilde{\eta}$,\cite{Mason1995Feb,Squires2009Dec,Puertas2014May,Waigh2005Feb} which enters the dynamic (stress relaxation) modulus via $\tilde{G}=-i\omega\tilde{\eta}$, yielding $\tilde{G}=\frac{m}{3\pi\sigma}(\omega^2-i\omega\tilde{K})$,
where $\text{Re}(\tilde{G})$  quantifies the elastic \emph{storage}, while the viscous \emph{loss} component is given by $\text{Im}(\tilde{G})$,\cite{BarratHansen,GraessleyBook,BirdBook,PokrovskiiBook,FerryBook1980} hence the latter is related to $\text{Re}(\tilde{K})$. For small $\omega$, the $\omega^2$-term can be neglected, yielding the simple scaling relation between $\tilde{K}$ and $\tilde{G}$, which allows for a  rough comparison with microrheology experiments.

 
%
%
%
%
%

\section{Analysis and Results}
\subsection{Tracer dynamical correlations}\label{sec:ACFana}
In \cref{fig:CvCfK} we depict the velocity autocorrelation $C_{v}(\tau)$ [panel (a)], the force autocorrelation  $C_{F}(\tau)$ [panel (b)], and the memory kernel $K(\tau)$ [panel (c)] as a function of the delay time $\tau$ for different tracer--network interaction strengths $\varepsilon\tb{tn}$ [color-coded, see legend in panel (a)], and two different polymer densities [see line pattern legend in panel (b)].

The decay time of velocity correlations [panel (a)] decreases with higher interactions strength and system density, and deviates significantly from the simple exponential decay $C_0(\tau)\propto\exp\left(-\gamma_0\tau\right)$ of the freely diffusing particle (gray line). The interactions with the dense surrounding thus reduce the mean free path and also the probability of the particles to perform free diffusion, hence reducing the effective diffusivity.

The interaction strength directly affects the magnitude of the deterministic forces acting on the tracer particles, as apparent in the force correlations [panel (b)]. High $\varepsilon\tb{tn}$ create a \emph{sticky} environment, i.e., tracers are likely to adsorb to polymer beads and pick up the network dynamics, evident by the emergent oscillations on the intermediate timescale, $\tau\gamma_0\approx 0.1$, in $C_v$ as well as in $C_F$. A more precise analysis of the frequencies and the source of such vibrations is provided later.

In the case of strong adsorption, the apparent decay in the velocity autocorrelation is increasingly determined by network oscillations and their damping times, and the density $\phi$ appears to have only little effect on the velocity and force correlations. We may conclude that the free diffusion barely occurs for large $\beta\varepsilon\tb{tn}$. The  motion is rather governed by hopping events along the vibrating polymer chains.

Note that the oscillations in $C_v$ and $C_F$ have the same frequencies, but are shifted by roughly one quarter of the oscillation period [compare the pronounced minima for strong interactions in \cref{fig:CvCfK}(a) and (b)]. This is expected due to the relation between force and velocity correlations, $C_F=-m^2\ddot{C}_v$, and can also be rationalized in the harmonic oscillator picture, in which the velocity lags behind the forces by ${\pi}/{2}$ due to the inertia $m$.

\subsection{\new{Tracer memory kernel}}\label{sec:kernelana}

The memory kernels $K$ presented in \cref{fig:CvCfK}(c) are related to $C_v$ and $C_F$. With $\tau\to0$, $K$ converges to \new{$(\beta/m)~C_F$}. Note that due to the logarithmic scaling of the delay time $\tau$, the delta-peak stemming from the solvent noise $\xi(t)$ in the microscopic model is not visible. Although the shape of $K$ at intermediate times, $\tau\gamma_0\approx0.1\text{-}1$,  significantly differs from $C_F$, we find again that the magnitude of the friction directly scales with the interaction strength $\varepsilon\tb{tn}$, as similarly observed for $C_F$, leading to a common intersection point of all presented kernels at  $\tau\gamma_0\approx0.1$. As already observed for $C_v$ and $C_F$, the increased dynamical coupling of the tracers with the network vibrations at higher $\varepsilon\tb{tn}$ is also obvious in the memory, which we discuss in more detail in \cref{sec:network_dyna,sec:freq_domain,sec:dampedharmonics}.

\begin{figure}[t]
 \includegraphics[width=\linewidth]{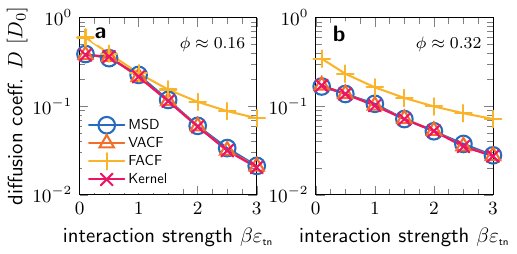}
 \caption{\label{fig:Dcompare} Long-time diffusion coefficients for different methods (cf. \cref{sec:diffusion_methods}) represented by different symbols and colors (see legend in panel (a)). The panels depict the results for different densities, (a) $\phi\approx0.16$ and (b) $\phi=0.32$. While the diffusion calculated from the MSD, VACF, and the kernel perfectly match, the FACF method only captures the rough scaling.}
\end{figure}

\subsection{Long-time diffusion}\label{sec:longtimediff}
In \cref{fig:Dcompare} we compare the effective diffusion coefficients obtained from the different methods, introduced in \cref{sec:diffusion_methods}, for seven different interactions strengths $\varepsilon\tb{tn}$ and two different polymer volume fractions $\phi$. The methods based on the MSD, VACF and the kernel represent the very same physics yielding the exact same diffusion coefficients, while the FACF estimate only reproduces the overall trend.

The essentially exponential scaling, $D\propto\exp(-A\beta\varepsilon\tb{tn})$, with scaling $A$, is known from similar coarse-grained simulations of diffusive tracers in flexible as well as in rather stiff polymer networks.\cite{Kim2020,Milster2021} At low interaction strength, volume exclusion effects $D\propto\exp(-C\phi/(1-\phi))$ dominate the diffusion process.\cite{Amsden1998Nov,Masaro1999Aug} In this regime, the higher densities reduce the probability for the particles to perform free diffusion, which is in line with the observations in the VACF [\cref{fig:CvCfK}(a)] and the kernel [\cref{fig:CvCfK}(b)]. Precisely, for $\beta\varepsilon\tb{tn}=0.1$, $C_v$ and $K$ are almost purely decaying functions, where $C_v$ decays faster with increasing $\phi$, and where $K$ shows increased magnitude.

Nonetheless, with increased interactions, we see that the scaling $A$ is smaller for $\phi\approx0.32$, and, eventually, the diffusion for tracers at very strong attractions ($\beta\varepsilon\tb{tn}=3$) is higher in the denser networks. This is usually rationalized by the overlaps of the numerous LJ potentials essentially smoothening the energy landscape and thus creating pathways through the network speeding up the diffusion.\cite{Kim2019Mar,Milster2021,Ghosh2015}

\subsection{Network dynamics}\label{sec:network_dyna}

\begin{figure}[b]
 \hspace*{-1cm}\includegraphics[scale=\sc]{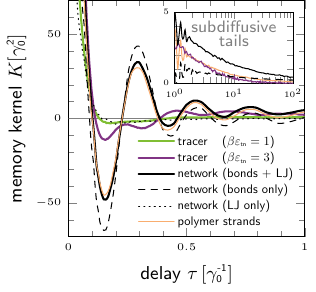}
 \caption{\label{fig:network_dyna}Comparison of the tracer memory kernel with the network memory kernel. Black solid line corresponds to the kernel averaged of all polymer beads with full force field (network--network LJ interactions $\beta\varepsilon=0.5 \Rightarrow \phi\approx0.32$  and bonded potentials) as used in the simulations (as in \cref{fig:CvCfK}). The black dashed line depicts $K$ for the same network at same density but without the LJ interactions, hence extracting pure bond contribution to the network memory. The black dotted line depicts the kernel of the polymer beads after the removal of all bonded potentials, hence representing a LJ gas with $\beta\varepsilon\tb{nn}=0.5$. \new{The solid orange line reflects the memory kernel of monomers in a polymer melt with strand length of ten monomers, no crosslinks, and $\phi\approx0.29$. The purple and the green solid lines (`tracer')} are the tracer memories from the original simulations for comparison (as in \cref{fig:CvCfK}). The inset depicts \new{four} memory kernels with clear subdiffusive regimes at large delay times. The long-time tail of the network memory kernel (with full force field) scales roughly with $\tau^{-1/3}$.  See main text for more details and interpretation. }
\end{figure}

The dynamical coupling between tracers and the network increases with stronger LJ interactions $\varepsilon\tb{tn}$. Analysing the memory kernel of the polymer beads thus reveals the origin of the observed oscillations in the tracer memory. In \cref{fig:network_dyna} we depict the single particle memory kernel of the (tracer-free) network \new{(with $\phi\approx0.32$)} averaged over all polymer beads (black solid line) in comparison with the tracer memory with moderate ($\beta\varepsilon\tb{tn}=1$) and very strong interactions ($\beta\varepsilon\tb{tn}=3$) in such a network [already presented in \cref{fig:CvCfK}(c)]. We also present the memory of the network particles for \new{three auxiliary simulations for which we disabled the non-bonded interactions (bond-only network), turned off all bonds (LJ gas), and removed all crosslinkers (melt of polymer strands with comparable density $\phi\approx0.29$), respectively.}

The first auxiliary simulation represents a network that only interacts via bond and angle potentials \new{(black dashed line in \cref{fig:network_dyna}) and exhibits the same underdamped oscillations apparent in the network with the full force field (black solid line).} We conclude that the vibrations due to the bonded potential dominate the polymer dynamics. The frequency $\omega\tb{bonds}\approx 25\gamma_0$ is of the order of a single bond oscillation, e.g., a single harmonic potential with spring constant $\kappa\tb{b}=100 \kB T / \sigma^2$ yields $\omega^*\tb{harm.}=\sqrt{2\kappa\tb{b}/m}\approx14\gamma_0$, and the two-bond stretch vibration gives $\omega^*\tb{stretch}=\sqrt{4\kappa\tb{b}/m}=20\gamma_0$.\cite{goychuk2012viscoelastic,tang2006chain} We also observe that the amplitude of the oscillation is more pronounced in the absence of LJ interactions, hence, the non-bonded interactions slightly inhibit the bond vibrations.

In the second auxiliary simulation \new{(black dotted line in \cref{fig:network_dyna})} all bonds were removed, resulting in a pure LJ system (with $\beta\varepsilon\tb{nn}=0.5$), and one observes the typical signature of a LJ gas with weak attractions.\cite{Shin2010Oct,jung2017iterative,Straube2020Jul} We will demonstrate later that the LJ contributions are described by strongly damped oscillation, with $\omega\tb{LJ}/\gamma_0\approx10\text{-}15$. In highly diluted systems, the pairwise frequency from a harmonic approximation at the LJ minimum \new{(at $r=2^{1/6}\sigma$)} provides the lower limit. It scales with $\omega^*\tb{LJ}\approx{\sigma^{-1}}\sqrt{25 \varepsilon / (2m)}$, which yields $\omega^*\tb{LJ}(\beta\varepsilon=0.5)=2.5\gamma_0$ in our case. The actual frequency is higher, since, on the one hand, the LJ minimum is considerably steeper than the second order approximation, and, on the other hand, due to collective effects, such as the superposition of multiple potential wells.

\new{Considering finally a polymer melt (11664 strands of ten monomers each and no crosslinks), which is our third auxiliary simulation (depicted as orange solid line in \cref{fig:network_dyna}), the memory kernel is almost identical to the one for network (with bonded and LJ interactions) at small timescales, attributed to the similarities in the local environment and interactions.}

\new{At large delay times we find long subdiffusive tails in the memory kernels for the `bonded networks' and the polymer strands} (see inset in \cref{fig:network_dyna}), while the LJ gas converges quickly to linear diffusion. In fact, we find strongest subdiffusion for the network with the full force field,  for which the tail of the memory kernel scales with roughly $\tau^{-1/3}$. Hence, the \new{very long} subdiffusion is a network effect, which is significantly magnified by LJ interactions.\cite{Kopf1997Nov,Kremer1990Apr} \new{This conclusion is further substantiated by the observed long-time memory of the polymer strands. While the magnitude of the subdiffusive tail is increased by the LJ interactions (compare with the bond-only network), it also decays considerable faster then the tails of both `bonded networks'.}

Inspecting the tracer memory, we find a clear LJ signature for the moderate interaction strength ($\beta\varepsilon\tb{tn}=1$), and only at higher interactions (e.g. $\beta\varepsilon\tb{tn}=3$) the tracers pick up the bond vibrations and also exhibit a subdiffusive regime. \new{ The subdiffusion of the tracers, however, decays notably faster than the ones of the `bonded networks', and the polymer strands.} Details on the dynamical coupling become evident in the frequency domain provided below.

\begin{figure}[h!]
 \includegraphics[scale=\sc]{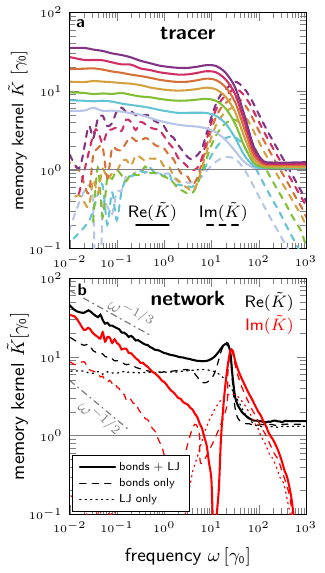}
 \caption{\label{fig:Komega}(a) Real and imaginary parts of the frequency-dependent tracer friction $\tilde{K}(\omega)$ in the dense network ($\phi\approx0.32$) for different interaction strengths $\beta\varepsilon\tb{tn}$, color-coded as in the \cref{fig:CvCfK} [from dark blue ($\beta\varepsilon=0.1$) to purple ($\beta\varepsilon=3.0$)]. The real part is decisive for viscous loss, where the low and high $\omega$ limits correspond to effective long-time friction $\tilde{K}(0)=\gamma\tb{kernel}$ and the uncorrelated solvent-related friction $\tilde{K}(\omega\to\infty)\approx\gamma_0$, respectively.  The imaginary part reflects elastic collisions with the polymer and rises with increasing coupling to bond vibrations ($\omega\tb{bonds}\approx25\gamma_0$), but also slower network modes in the low-frequency domain ($\omega/\gamma_0\approx0.1$-$1$). For $\omega\to0$, we observe a vanishing imaginary part, indicating that the adsorbed particles eventually detach from the polymer for very long times. (b) Real (black) and imaginary (red) parts of network memory in the frequency domain for the network ($\phi\approx0.32$) with the full force field and the two auxiliary simulations, the bond-only network and the LJ gas. The low-frequency scaling $\text{Re}(\tilde{K})\propto\text{Im}(\tilde{K})\propto\omega^{n-1}$ with $n-1\approx-1/3$ can be computed via \new{$n\approx\left(2/\pi\right)\arctan[-\text{Re}(\tilde{K})/\text{Im}(\tilde{K})]$}, which is related to the dynamic relaxation modulus $\tilde{G}(\omega)$ of the whole network.}
\end{figure}

\subsection{Frequency domain}\label{sec:freq_domain}
Analyzing the memory in the Fourier space allows us to identify the spectral density of the correlated fluctuations $\boldsymbol{\zeta}$, interpret the frequency-dependent friction, and extract the dominant vibration modes contributing to the \new{memory kernel}. In  \cref{fig:Komega}(a) we depict $\tilde{K}(\omega)$  for different $\varepsilon\tb{tn}$ and for tracers in the dense network only. The low-frequency limit, $\text{Re}[\tilde{K}(0)]=\gamma\tb{kernel}$, is identical with the integral of the memory kernel [\cref{eq:einstein_relation}], and confirms the above presented long-time limit. At the low, yet non-zero frequencies, the slow convergence of $\text{Re}[\tilde{K}(\omega)]$ to $\gamma\tb{kernel}$ signifies the existence of very long subdiffusive cross-over regimes, corresponding to the long tails in $K(\tau)$ .

At high frequencies $\text{Re}[\tilde{K}(\omega)]$ shows low-pass characteristics, but the friction is limited by uncorrelated solvent friction $\gamma_0$ (indicated by gray horizontal line). Note that $\gamma_0$ is not perfectly reached since the polymer beads are also subject to uncorrelated solvent fluctuations, which are passed on to the tracers with increasing coupling strengths $\beta\varepsilon\tb{tn}.$

At intermediate frequencies ($\omega/\gamma_0\approx10\text{-}50$) we find the superposition of several damped oscillations, whose magnitudes and resonating behaviors increase with $\varepsilon\tb{tn}$. The maximization of the imaginary part indicates the presence of elastic interactions and corresponds to the emergent oscillations observed in the correlation functions and in the memory kernel (\cref{fig:CvCfK}). The  plateau in the imaginary part at lower frequencies originates from slower, elastic network deformations [cf. network memory in frequency domain in \cref{fig:Komega}(b)], yet only plays a minor role due to the overall large friction in the low frequency regime. The real part, in contrast, is dominant and approaches slowly to the long-time limit ($\omega\to0$), since the adsorbed tracers adopt the network subdiffusion for relatively long timescales $\omega\approx0.1\gamma_0$, and only detach from the polymer at very long times, which also coincides with the vanishing $\text{Im}(\tilde{K})$.

Inspecting the network kernel and the auxiliary simulations, depicted in \cref{fig:Komega}(b), allows us to deconvolute the effects again. We see a strong resonating behavior stemming from the bond vibrations, a rather damped signature from the LJ interactions, and low-frequency scaling typical for networks. The latter can, to some extend, be compared to viscoelastic experiments of gels. For instance, Winter\cite{Winter1987Dec} evaluated the frequency scalings of the storage and loss modulus during crosslinking polymerization in order to detect the gelation point (fully connected polymer network) and proposed the general scaling $\text{Re}(G)\propto\omega^n$ and $\text{Im}(G)\propto\omega^n$ with $0<n<1$, which relates the storage and loss, reading $\tan(n\pi/2)=\text{Im}(G)/\text{Re}(G)$.
Given the GSER is valid in the low-frequency domain, one may assume \new{$n\approx\left(2/\pi\right)\arctan[-\text{Re}(\tilde{K})/\text{Im}(\tilde{K})]$}, resulting in $n\approx2/3$ for $\omega\lesssim0.1$ in our simulations, well in line with typical hydrogels ($n\approx0.5$ - $0.8$).\cite{Kavanagh1998Jan} 

This further provides the correct scaling ($n-1$) for the memory kernel, i.e.,  $\text{Re}(\tilde{K})\propto\omega^{-1/3}$ and $\text{Im}(\tilde{K})\propto\omega^{-1/3}$, as depicted in \cref{fig:Komega}(b), reflecting the $\tau^{-1/3}$-scaling of the subdiffusive tails (cf. inset in \cref{fig:network_dyna}). We hence may conclude that the low-frequency scaling is dominated by very slow network dynamics. In fact, it decouples completely from the fast bond vibrations, evident in the comparison of $\tilde{K}$ for networks with varying bond stiffness (see. \cref{fig:bond_compare} in the Appendix \labelcref{app:network_memory}).




\begin{table}[b]
\caption{\label{tab:fitN4}Memory kernel fit parameters for $\phi\approx0.32$ and attractive tracer--network interactions $\varepsilon\tb{tn}\in\{1.0,1.5,2.0,2.5,3.0\}$ with $N_k=4$ modes. The kernels with different $\varepsilon\tb{tn}$ were fit simultaneously by \cref{eq:kernelfit} and introducing the scaling [cf. \cref{eq:epsscaling}] with parameter $\alpha_k$. The delta peak stemming from the Langevin thermostat [cf. \cref{eq:CLE}]  was not included for fitting and needs to be added manually to the memory kernel to account for the solvent-related friction.  }
 \begin{tabular}{S[table-format=1] S[table-format=3.1]  S[table-format=1.2]  S[table-format=2.1]  S[table-format=1.2] l }
  &{ \sffamily magnitude } & { \sffamily decay }  & { \sffamily frequency } & \sffamily{  scaling }& { \sffamily origin }\\
   {$k$} &{$\lambda_k~[\gamma^2_0]$} &{$\tau_k~[\gamma_0^{-1}]$}& {$\omega_k~[\gamma_0]$} & {$\alpha_k$} & \\\hline
  1 & 118.6 & 0.08 & 10.4 & 0.94 & \sffamily LJ\\
  2 & 38.0  & 0.08 & 25.5 & 1.29 & \sffamily bonds\\
  3 & 4.5   & 0.08 & 43.8 & 1.64 & \\
  4 & 0.6   & 2.49 & 0.0  & 1.73 & \sffamily network\\
  {(0)} & \multicolumn{4}{c}{\sffamily delta peak at $\tau=0$} & \sffamily solvent
 \end{tabular}
\end{table}

\subsection{Damped harmonics approximation}\label{sec:dampedharmonics}
We now approximate the memory kernel with a series of damped oscillations, reading
\EQ{
K(\tau)=\sum^{N_k}_{k=1}\lambda_k\exp\left(-\frac{\tau}{\tau_k}\right)\cos\left(\omega_k\tau\right),\label{eq:kernelfit}
}
summarizing the leading modes and their dependence on tracer--network coupling $\varepsilon\tb{tn}$. In fact, the magnitude for constant $\phi$ scales with $\varepsilon\tb{tn}$, and we extend \cref{eq:kernelfit} with
\EQ{
\lambda_k\to\lambda_k (\beta\varepsilon\tb{tn})^{\alpha_k},\label{eq:epsscaling}
}
where $\alpha_k$ is the mode-dependent scaling. \cref{eq:kernelfit,eq:epsscaling}  are suitable for interpolation of the memory kernel, i.e., $K(\varepsilon\tb{tn};\tau)$, and can be used to simulate the GLE with auxiliary variables (Markovian embedding).\cite{Shugard1977Mar,Li2017Jan,Wang2020,Brunig2022Feb,Doerries2021Mar}

With this extension we aim to fit as many memory kernels simultaneously as possible with the smallest number of modes. For demonstration, we use the dense system ($\phi\approx0.32$) and only attractive interactions ($\beta\varepsilon\tb{tn}\in\{1.0,1.5,2.0,2.5,3.0\}$), i.e., the regime in which tracer adsorption to the polymer plays a significant role. The fitting focuses on the short and intermediate timescales and only considers the kernel data up to $\tau=10\gamma_0^{-1}$. The fitting results with $N_k=4$ modes are summarized in \cref{tab:fitN4}, ordered by magnitude $\lambda_k$, and confirm our observations from the previous sections.

\begin{figure}[t]
 \includegraphics[width=\linewidth]{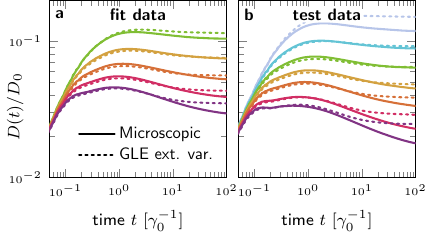}
 \caption{\label{fig:MSDcompare}Comparison of $D(t)=\text{MSD}/(6t)=\int_{0}^tC_v(t')\mathrm{d}t'$ for microscopic simulations and GLE simulations based on the Markovian embedding (see Appendix \labelcref{appendix_GLE} for details) for the (training) fit data (dense attractive networks: $\phi\approx0.32$, $\beta\varepsilon\tb{tn}\in\left\{1.0, 1.5, 2.0, 2.5, 3.0\right\}$) [panel (a)] and some test data ( $\beta\varepsilon\tb{tn}\in\left\{0.75, 1.25, 1.75, 2.25, 2.75, 3.5, 4.0\right\}$) [panel (b)], which was not considered for the fit.  The fitting parameters provided in \cref{tab:fitN4} were obtained by fitting the kernels of the fit data simultaneously based on \cref{eq:kernelfit} and the $\varepsilon\tb{tn}$-scaling \cref{eq:epsscaling}, which were used to inter- and extrapolate kernels for the GLE simulations for the test data.    }
\end{figure}

{The dominant contribution ($k=1$) at small delay times stem from oscillations close to critical damping with a frequency of $\omega_1\approx 10 \gamma_0$. We have already identified this mode as a clear LJ signature, and the almost linear scaling ($\alpha_1\approx1$) with $\beta\varepsilon\tb{tn}$ confirms this interpretation. The second strongest mode with $\omega_2\approx25\gamma_0$ is most obvious at intermediate delay times, $\tau\gamma_0\approx0.1\text{-}1.0$, and quantifies the coupling of the tracer dynamics to the polymer bond vibrations with a scaling of $\alpha_2\approx1.3$. The third mode has rather small impact and cannot be directly attributed to a single interaction type. It is rather a result of the assumption that only the magnitude scales with the LJ interactions. In fact, this mode shares the same decay time as the dominating ones, and due to $\omega_3\approx44\gamma_0$ it essentially performs a small shift of the extrema to smaller $\tau$ with increasing $\varepsilon\tb{tn}$, indicating that also $\omega_k$ and $\tau_k$ for the first two modes slightly scale with $\varepsilon\tb{tn}$. Mode four has the smallest magnitude and only plays a minor role for small $\tau$. However, it has a significantly longer decay of $\tau_4\gamma_0\approx 2.5$ and is decisive for the long subdiffusive tails observed for strong LJ attractions.

We verify our fitting approach by comparing the time-dependent diffusion coefficients, $D(t)=\int_{0}^tC_v(t')\mathrm{d}t'$, of the microscopic simulations with the GLE simulations presented in \cref{fig:MSDcompare}. Details on the GLE simulations are provided in the Appendix \labelcref{appendix_GLE}. In panel (a) we compare the two simulation techniques for the fit data and only observe significant deviations for larger times and very strong coupling. For the results depicted in panel (b) we have inter- and extrapolated the GLE model and compare the results to those of microscopic simulations that were not used for fitting. We find a very good short-time agreement up to $t\approx\gamma_0^{-1}$. At relatively long timescales, up to $t\approx\gamma_0^{-1}$, the interpolated kernels perform well, and only at very large times we see significant deviations, particularly for extrapolated kernels \new{at the largest and smallest tested $\varepsilon\tb{tn}$ values.}

\section{Concluding remarks}

\new{As motivated in the introduction, tracer diffusion in polymer networks is not only important for applications but provides valuable insight into the physics of fluctuations and transport in soft, viscoelastic media. Long-time relaxation modes, describable by memory kernels, are key to the rationalization of the viscoelasticity of the polymer matrix but a systematic study of the kernel behavior depending on the tracer--matrix coupling had been missing, yet. Here, we presented a well-controlled tracer--network simulation model, in which memory kernels could be calculated with unprecedented statistical quality. This allowed us to identify various, frequency-dependent contributions of the kernel, and to discuss the origins of the emerging physics separately, depending on the tracer--polymer coupling.


In particular, with increasing attraction strength between tracer and polymer, the tracer is more likely to adhere to the polymer matrix and pick up its dynamics; in here, we identified the signature of the LJ liquid, emergent bond vibrations, and the slow subdiffusion of the whole network on larger scales. The magnitude of these effects entering the kernel scales almost linear with the LJ interaction strength. With that we demonstrated how the single solute kernel analysis may be employed to retrieve dynamic and viscoelastic information about the solute transport as well as about the surrounding polymeric matrix}.

\new{The analysis allowed us further to construct a GLE in which the polymer network enters effectively only through the essential modes (using an auxiliary variable method) in the memory kernel and the corresponding fluctuations.  In this reduced description, the polymer network serves now as an additional bath with controlled coupling to the tracer system, enabling the efficient inter- and extrapolation of the friction effects.  This will be useful to predict the outcome of other couplings and modified system setups in efficient GLE-based simulations in future.

The recognized modes may also be useful to construct or optimize empirical kernels to describe and interpret specific experiments, e.g., for the mechanical tracer response in  in the strong tracer--polymer coupling limit,\cite{Rizzi2020Jul,Grebenkov2013Oct,goychuk2012viscoelastic} or for the nonequilibrium response in biological networks, as demonstrated recently.\cite{PhysRevLett.131.228202} Other situations for extrapolation and applications are relatively large tracers (larger than the typical polymer mesh sizes) which lead to certain collective deformations of the underlying elastic network,\cite{PhysRevX.4.021002} of interest in the realm of microrheology.\cite{Winter1987Dec, Waigh2005Feb,Tassieri2012Nov, Robertson-Anderson2018Aug}}

Finally, our framework, developed under homogeneous equilibrium conditions, serves as a solid basis to extend the dynamical coarse-graining procedure to inhomogeneous setups or to nonstationary, nonequilibrium situations in future,\cite{schilling2022, Netz2023} e.g., for externally driven tracer systems,\cite{Q:nonlinear} possibly including time-delayed feedback.~\cite{feedback,Gernert2015Aug,Loos2019Feb} \new{Importantly, adding external forces to the underlying microscopic model may result in an additional terms describing \emph{nonequilibrium response forces} in the GLE, which can only be neglected in the zero-force limit.}\cite{Koch2023Sep}

\section*{Acknowledgments}
The authors thank Roland R. Netz for fruitful discussions. This work was supported by the Deutsche Forschungsgemeinschaft (DFG) via the Research Unit FOR 5099 ``Reducing complexity of nonequilibrium systems'' and Project No. 430195928.  The authors also acknowledge support by the state of Baden-Württemberg through bwHPC and the DFG through grant no INST 39/963-1 FUGG (bwForCluster NEMO) and under Germany's Excellence Strategy - EXC-2193/1 - 390951807 ('LivMatS').

\section*{Author declarations}

\subsection*{Conflict of Interest}
The authors have no conflicts to disclose.


 \section*{Data Availability}
The  data  that  support  the  findings  of  this  study  are  available from the corresponding author upon reasonable request.

\section*{Appendix}
\appendix

\begin{table}[b]
	\caption{\label{tab:simdetails}Parameter summary for the microscopic model. }
	\begin{tabular}{l l l}\sffamily
		\sffamily total polymer number &   & 122472 \\
		\sffamily chain monomer number & $N\tb{mer}$ & 116640 \\
		\sffamily crosslinker number & $N\tb{x}$ & 5832\\
		\sffamily crosslink ratio & $N\tb{x}/N\tb{mer}$ & $5\%$\\
		\sffamily tracer concentration & $c$ & $0.005\sigma^{-3}$\\[1em]
		\sffamily thermal energy & $\kB T$ & \sffamily \newx{unit}\\
		\sffamily bead diameter & $\sigma$ & \sffamily \newx{unit}\\
        \sffamily \newx{bead mass} & \newx{$m$} & \sffamily \newx{unit} \\
		\sffamily solvent friction & $\gamma_0$ & \sffamily \newx{unit} \\
		\sffamily \newx{free diffusion} & \newx{$D_0$} & \sffamily \newx{$\kB T/(m\gamma_0)$} \\
		\sffamily Brownian timescale & $\tau_0$ & \sffamily $\gamma_0^{-1}$\\[1em]
		\sffamily bond stiffness & $\kappa\tb{b}$ & $100~\kB T/\sigma^2$\\
		\sffamily eq. bond length & $r_0$ & $\sigma$ \\
		\sffamily angle stiffness & $\kappa\tb{a}$ & $1~\kB T/\text{rad}^2$ \\
		\sffamily eq. angle (chain) & $\Psi\tu{0}\tb{mer}$ & $\pi$\\
		\sffamily eq. angle (crosslink) & $\Psi\tu{0}\tb{x}$ & $\arccos(-1/3)$ \\[1em]
		\sffamily tracer--network interaction & $\varepsilon\tb{tn}$ & $0.1~\kB T$\\
		\sffamily tracer--tracer interaction & $\varepsilon\tb{tt}$ & $\left[0.1, 3.0\right]~\kB T$ \\
		\sffamily network--network interaction & $\varepsilon\tb{nn}$ & $\left\{0.35, 0.50\right\}~\kB T$\\
		\sffamily polymer volume fraction & $\phi$ & $\left\{0.16, 0.32\right\}~\kB T$\\
	\end{tabular}
\end{table}

\section{Microscopic simulation details}\label{app:mic_sim}{
The creation of the regular polymer network and the simulations of the Markovian Langevin model, \cref{eq:CLE}, were performed with the LAMMPS\cite{LAMMPS} software package, similar to our previous work.\cite{Milster2021} The network consist of 122472 polymer beads and has a crosslink ratio of 5\% yielding a chain length of ten monomers between two crosslinker beads. We fixed the bond and angle potentials (see \cref{tab:simdetails}) and tuned the polymer volume fraction $\phi$ with the LJ network--network interactions $\varepsilon\tb{nn}$ by equilibrating the network in $NpT$ ensemble with an isotropic Berendsen barostat ($\tau_p=0.1\tau_0$,~$p\tb{target}=0$) and the Langevin integration scheme. The equilibration was carried out until convergence of the cubic simulation box volume was reached, yielding $\phi\approx0.16$ for $\beta\varepsilon\tb{nn}=0.35$ and
$\phi\approx0.32$ for $\beta\varepsilon\tb{nn}=0.5$. This corresponds to box lengths of $L=73.3\sigma$ and $L=59.5\sigma$ for the moderate and the high polymer density, respectively.

After the network equilibration, the simulation box volume was fixed, and 1956 (for $\phi\approx0.16$) and 1041 tracer particles (for $\phi\approx0.32$) were randomly added to the system, yielding a tracer concentration of $c\approx0.005\sigma^{-3}$. The tracer positions were subsequently minimized and the whole system was again equilibrated in $NVT$  ensemble (Langevin integration) for $\tau\tb{EQ}=100\tau_0$ with a timestep of $\mathrm{d}t=0.001\tau_0$. The production run was carried out with identical timestep for $\tau\tb{PR}=10^3\tau_0$ and the tracer trajectories were output every timestep.

In order to evaluate the dynamics of the network beads we ran additional simulations without tracer solutes. Instead of storing the single particle trajectories, we made use of LAMMPS' on-the-fly calculation of the velocity autocorrelation function (VACF), providing the VACF averaged over all network beads. We performed 40 consecutive production runs for $\tau\tb{PR}=10^3\tau_0$ with $\mathrm{d}t=0.005\tau_0$, yielding 40 individual VACFs, where the initial timestep of each run was used as reference for the delay. The final result was obtained by averaging over these individual VACFs. This procedures was employed for the network with the full force field as well as for the auxiliary simulations with partial force fields [as presented in \cref{fig:network_dyna} and \cref{fig:Komega}(b)]. From the VACFs the memory kernel was computed as described in \cref{sec:computeK}.
}

\section{The Generalized Langevin Equation}\label{appendix_GLE}
In this appendix we briefly recall the main results of Mori's projection operator formalism.\cite{mori1965transport} Starting from the (Hamiltonian) equations of motion of the microscopic system, the projection operator formalism together with the linear Mori projector yields in a stationary distribution the following coarse-grained equations of motion 
\begin{align}
	m \dot{v}_{i\nu}(t) &= am v_{i\nu}(t) - m\int\limits_0^t\dd t'\,K(t-t')v_{i\nu}(t')+\zeta_{i\nu}(t).
\end{align}
Here, the different quantities can be expressed as
\begin{align}
	a&=\frac{(\mathcal{L}v_{i\nu}, v_{i\nu})}{(v_{i\nu},v_{i\nu})},\\
	K(t) &=- \frac{\left(\mathcal{L}\exp(t\mathcal{Q}_\text{M}\mathcal{L})\mathcal{Q}_\text{M}\mathcal{L}v_{i\nu},v_{i\nu}\right)}{(v_{i\nu},v_{i\nu})},\\
	\zeta_{i\nu}(t) &= m\exp(t\mathcal{Q}_\text{M}\mathcal{L})\mathcal{Q}_\text{M}\mathcal{L}v_{i\nu},
\end{align}
where $(X,Y)$ denotes the inner product,
\begin{align}
	(X, Y) &= \int\dd\boldsymbol{\Gamma} \rho_0(\boldsymbol{\Gamma}) X(\boldsymbol{\Gamma})Y(\boldsymbol{\Gamma}),\label{eq:def_inner_product}
\end{align}
and $\mathcal{Q}_\text{M}$ is the orthogonal Mori projection operator defined by its action on an arbitrary phase-space function,
\begin{align}
	\mathcal{Q}_\text{M}X(\boldsymbol{\Gamma}) &= X(\boldsymbol{\Gamma}) - \frac{(X,v_{i\nu})}{(v_{i\nu},v_{i\nu})}v_{i\nu}. 
\end{align}
The $\rho_0(\boldsymbol{\Gamma})$ in \cref{eq:def_inner_product} denotes the stationary phase-space distribution sampled from the simulations or experiments. From these definitions it is easy to see that the Liouvillian is an anti-self-adjoint operator, i.e., $a=0$, and that the fluctuation-dissipation theorem
\begin{align}
	\left\langle \zeta_{i\nu}(t)\zeta_{i\nu}(t')\right\rangle &= m^2 K(t-t')\left\langle v_{i\nu}^2\right\rangle
\end{align}
holds true.\\
Note that the entire formalism can also be used for simulations that involve a Langevin thermostat as shown in ref.~\citenum{Glatzel_2021_Comments}.

\section{Numerical details on the kernel calculation}\label{appendix_numerics_of_memory_kernel}
Here, we briefly describe the numerical algorithm and refer to ref.~\citenum{Meyer_2020_Numerical} for the details of the derivation, and use a notation close to the one used therein to make a comparison with the original method as easy as possible. The starting point is the sampled time-discrete correlation function $C_v(n\,\Delta t)$ for $n\in \{0,\ldots,N-1\}$. A numerical differentiation together with a simple scaling yields $S_0(n\,\Delta t):=-\dot{C}_v(n\,\Delta t) / C_v(0)$. We set $S(0)=1$ and calculate $S(n\,\Delta t)$ via
\begin{align}
    S(n\,\Delta t) &= \Delta t\sum\limits_{i=0}^{n-1} S_0((n-1-i)\Delta t) S(i\,\Delta t).
\end{align}
Subsequently, we numerically calculate the convolution of $S_0(n\,\Delta t)$ and $S(n\,\Delta t)$, reading
\begin{align}
    J(n\,\Delta t) &:= \sum\limits_{i=0}^n S_0(i\,\Delta t) S((N-1-i)\Delta t),
\end{align}
and obtain the memory kernel via a final numerical differentiation $K(n\,\Delta t)=\dot{J}(n\,\Delta t)$.

\begin{figure}[t]
\includegraphics[scale=\sc]{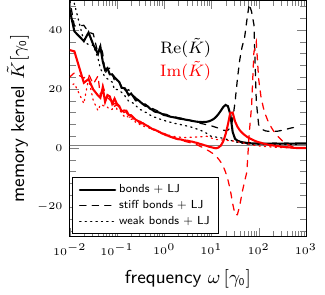}
\caption{\label{fig:bond_compare} Network memory in frequency domain for the network (with full force field) and different values of the bond stiffness $\kappa\tb{b}\in\{10, 100, 1000\} \kB T/\sigma^2$. While the frequency ($\omega\tb{b}\propto\sqrt{\kappa\tb{b}}$) and magnitude of the fast bond vibrations increases with $\kappa\tb{b}$, the low frequency network relaxation is independent of the bond stiffness, and is contributed to the mere network constraints, enhanced by LJ interactions (cf. \cref{fig:Komega}) }
\end{figure}

\section{Network memory in frequency domain}\label{app:network_memory}
To demonstrate that the long-time scaling (low frequency domain) observed in the polymer kernel is a whole-network effect, and thus independent of the bond strengths, we performed additional auxiliary simulations with differing bond strengths, $\kappa\tb{b}\in\{10, 100, 1000\} \kB T/\sigma^2$. As obvious in \cref{fig:bond_compare}, the shifted resonance signatures at higher frequencies are in line with the expected bond vibrations, while the low-frequency scaling  ($\approx\omega^{-1/3}$) is not affected by $\kappa\tb{b}$.

\section{GLE simulations}
The GLE can be integrated directly using the stored trajectory and the memory kernel. While this method is very efficient for rapidly decaying kernels, it becomes very costly for long memories. However, it has been shown that one can rewrite one non-Markovian GLE into a set of Markovian Langevin equations with auxiliary/extended variables,\cite{Shugard1977Mar,Li2017Jan,Wang2020,Brunig2022Feb,Doerries2021Mar} given one can approximate the kernel with damped oscillations. This method originates from the fact that the memory kernel can be calculated analytical for a single particle that is coupled to individual Brownian oscillators. Note that in a precise derivation,\cite{Shugard1977Mar} the resulting series has a cosine and an additional sine term, i.e., $K=\sum_k\lambda_k\exp(\tau/\tau_k)[\cos(\omega_k \tau)+\omega_k^{-1}\tau_k^{-1}\sin(\omega_k \tau)]$, while we disregard the latter for simplicity in \cref{eq:kernelfit}. In fact, we aim for a rather simple and amenable coarse-grained description capturing the dominant dynamics with only a few leading modes, for which an additional sine term is known to be practically irrelevant.\cite{Li2017Jan} The major deviation between the microscopic simulations and our GLE simulations (see \cref{fig:MSDcompare}) are dominated by the imperfect global fitting (only four modes for five kernels) and its scaling approximation $\lambda_k(\beta\varepsilon\tb{tn})^{\alpha_k}$.

The Markovian equivalent to the GLE for a single tracer in one dimension reads

\EQ{
\dot x &= v,\\
m \dot v &= -  m\sum_k \mathcal{A}_k~(s^k_1+s_2^k),\label{eq:extvarV}
}
where the auxiliary variables, $s_1^k$ and $s_2^k$, have the units of velocity, and model the friction and the correlated fluctuations simultaneously. Each mode $k$ is governed by a set of linear stochastic differential equations,
\EQ{
\dot s_1^k&=\mathcal{A}_kv  - \mathcal{B}_ks_1^k -\mathcal{C}_k s_2^k +\xi_k(t),\\
\dot s_2^k&=\mathcal{A}_kv +\mathcal{C}_ks_2^k.
}
Due to neglecting the sine term, the coefficients take a rather simple form, i.e., $\mathcal{A}_k=\frac{1}{4}\sqrt{\lambda_k}$, $\mathcal{B}_k=2\tau_k^{-1}$, $\mathcal{C}_k=(\omega_k^2+\tau_k^{-2})^{1/2}$. Each mode has delta-correlated noise $\langle\xi_k(t)\xi_k(t')\rangle=\kB T 4\tau_k^{-2}\delta(t-t')$, satisfying the FDT.

To account for the solvent-related fluctuations, we added $-m\overline{\gamma}_0v+\overline{\xi}_0(t)$ with $\langle \overline{\xi}_0(t)\overline{\xi}_0(t')\rangle=2\kB T m \overline{\gamma}_0\delta(t-t')$  to \cref{eq:extvarV}, where $\overline{\gamma}_0$ actually also scales with $\varepsilon\tb{tn}$ [see high frequency limit in \cref{fig:Komega}(b)], but we approximated it by $\overline{\gamma}_0={\gamma}_0$ for simplicity without noticeable impact. In fact, even dropping the solvent-related terms completely would not significantly change the simulation results (not presented) since the tracer dynamics are dominated by the polymer bath in the attractive regime. 

\new{
\section{List of Acronyms}
\begin{table}[h!]\caption{\new{List of acronyms used in this article.}}
\new{\begin{tabular}{l l}
LJ & Lennard-Jones\\
VACF & velocity autocorrelation function\\
FACF & force autocorrelation function\\
GLE & generalized Langevin equation \\
FDT & fluctuation--dissipation theorem \\
MSD & mean squared displacement  \\
GSER & generalized Stokes-Einstein relation\\[0.1cm]
$NVT$ & canonical ensemble:\\
&const. particle number $N$, volume. $V$, and temperature $T$\\[0.1cm]
$NpT$ & isothermal–isobaric ensemble:\\
& const. particle number $N$, pressure $p$, and temperature $T$
\end{tabular}}
\end{table}

}
\section*{References}

\bibliography{bib}

\end{document}